\documentclass[12pt]{article}
\usepackage[dvips]{color,graphicx}
\usepackage{hyperref}
\usepackage{amsmath,amssymb}
\usepackage{epsfig}
\usepackage[english]{babel}
\usepackage{pgf,pgfarrows,pgfnodes,pgfautomata,pgfheaps}
\usepackage{amsfonts,amsmath,amsthm,amssymb,enumerate,array,amssymb}
\usepackage[latin1]{inputenc}
\usepackage{bm}
\usepackage{fixmath}
\usepackage{amsbsy}
\usepackage{graphicx}
\usepackage{capt-of}
\usepackage{dcolumn}
\usepackage{yfonts}

\newcommand{\ap}{\ensuremath{\alpha'}} % Inverse string tension
\def\p{\partial}

\newcommand{\cL}{\mathcal{L}}

\newcommand{\cN}{{\mathcal{N}}}
\newcommand{\cO}{{\mathcal{O}}}

\newcommand{\be}{\begin{equation}}
\newcommand{\ee}{\end{equation}}
\newcommand{\bea}{\begin{eqnarray}}
\newcommand{\eea}{\end{eqnarray}}

\begin{document}

\begin{titlepage}

\begin{flushright}
UTTG-14-12\\
TCC-014-12
\end{flushright}
\vspace{0.3cm}
\begin{center} \Large \bf Holographic Brownian Motion in\\
Magnetic Environments 
\end{center}
\vspace{0.1cm}
% \vskip 0.1cm
\begin{center}
Willy Fischler\footnote{fischler@physics.utexas.edu},
Juan F. Pedraza\footnote{jpedraza@physics.utexas.edu}
and Walter Tangarife Garcia\footnote{wtang@physics.utexas.edu}

\vspace{0.2cm}
Theory Group, Department of Physics and Texas Cosmology Center, \\
The University of Texas, 2515 Speedway Stop C1608, Austin, TX 78712-1197\\
\vspace{0.7cm}
\end{center}

\begin{center}
{\bf Abstract}
\end{center}
\noindent
Using the gauge/gravity correspondence, we study the dynamics of a heavy quark in two strongly-coupled systems at finite temperature: Super-Yang-Mills in the presence of a magnetic field and non-commutative Super-Yang-Mills. In the former, our results agree qualitatively with the expected behavior from weakly-coupled theories. In the latter, we propose a Langevin equation that accounts for the effects of non-commutativity and we find new interesting features. The equation resembles the structure of Brownian motion in the presence of a magnetic field and implies that the fluctuations along non-commutative directions are correlated. Moreover, our results show that the viscosity is smaller than the commutative case and that the diffusion properties of the quark are unaffected by non-commutativity. Finally, we compute the random force autocorrelator and verify that the fluctuation-dissipation theorem holds in the presence of non-commutativity.

\vspace{0.2in}
\smallskip
\end{titlepage}
\setcounter{footnote}{0}

\tableofcontents

\section{Introduction}

Over the years, the study of thermal properties in quantum field theories has been of great interest. In the last decade, the discovery of the gauge/gravity correspondence \cite{Maldacena:1997re,Gubser:1998bc,Witten:1998qj} has provided tools for the study of a large class of strongly-coupled non-abelian gauge theories. Aside from pure theoretical motivation, one the the main reasons for pursuing such studies is the possible connection with the phenomenology of ultra-relativistic heavy ion collisions and/or strongly correlated condensed matter systems. An excellent account of these efforts can be found in the reviews \cite{Gubser:2009md,CasalderreySolana:2011us,Hartnoll:2009sz,McGreevy:2009xe} and the references therein.

The simplest and best-understood example of the correspondence relates four-dimensional $\cN=4$ Super-Yang-Mills (SYM) to type-IIB string theory (or supergravity) in asymptotically AdS$_5 \times {\rm S}^5$. In particular, the inclusion of a black hole in the bulk is known to be dual to a strongly-coupled gauge theory at finite temperature. The prescription to compute correlation functions using the gauge/gravity duality was developed a decade ago in \cite{Gubser:1998bc,Witten:1998qj} and was subsequently generalized to the real-time finite-temperature formalism in \cite{Son:2002sd,Herzog:2002pc}. This tool opened the possibility of computing the decay rates and timescales for the approach to thermal equilibrium of certain disturbances \cite{Horowitz:1999jd,Birmingham:2001pj,Kovtun:2005ev}, which in the gravity side are mapped to the computation of quasinormal frequencies. Perhaps, the most intriguing result in this front has been the well-known result of $1/4\pi$ for the shear viscosity to entropy density ratio at strong coupling \cite{Policastro:2001yc,Kovtun:2004de} which in turn led to the holographic calculation of many other transport coefficients \cite{Son:2007vk,Rangamani:2009xk}. More recently, it has been possible to gain further insight in the previously uncharted out-of-equilibrium regime by considering geometries that evolve in time \cite{Hubeny:2010ry,Balasubramanian:2010ce,Balasubramanian:2011ur,Caceres:2012em}.

In addition, other thermal properties have been inferred by considering various types of partonic
probes, and analyzing the manner in which the plasma damps
their motion. These studies include quarks \cite{Herzog:2006gh,Gubser:2006bz,CasalderreySolana:2006rq}, mesons \cite{Peeters:2006iu,Liu:2006nn,Chernicoff:2006hi}, baryons \cite{Chernicoff:2006yp,Krishnan:2008rs}, gluons \cite{Chernicoff:2006yp,Gubser:2008as} k-quarks \cite{Chernicoff:2006yp} and various types of defects \cite{Janiszewski:2011ue}. Although all these probes provide different information regarding the nature of the plasma, in this paper we focus on the dynamics of heavy quarks and their interactions with the thermal bath.

In the context of this duality, a heavy quark on the boundary theory corresponds to the endpoint of an open string that, at finite temperature, stretches between the boundary and the black hole horizon. The seminal works \cite{Herzog:2006gh,Gubser:2006bz} focused on the energy loss of a quark that is either moving with constant velocity as a result of being pulled by an external force, or is unforced but moving nonrelativistically and
about to come to rest. Further analyses \cite{Gubser:2006nz,CasalderreySolana:2007qw,Chernicoff:2008sa} made it clear that this mechanism of energy loss is closely related with the appearance of a worldsheet horizon (not to be confused with the spacetime horizon).

The study of quark fluctuations due to its interaction with the thermal bath involves going beyond the classical description of the string. As customary,
small perturbations about the average embedding are described
by free scalar fields propagating on the corresponding induced worldsheet
geometry. These fields can be excited due to Hawking radiation emitted by the worldsheet horizon, which in turn populates the various
modes of oscillation of the string. It was then found that, once these modes are quantized, the induced motion of the string endpoint is correctly described in terms of Brownian motion and its associated Langevin equation \cite{brownian,Son:2009vu}. This result was obtained by two different approaches:
the authors of \cite{brownian} reached this conclusion by assuming (following \cite{Lawrence:1993sg,Frolov:2000kx}) that the state of the quantized fields is the usual Hartle-Hawking vacuum, which describes the black
hole in equilibrium with its own thermal radiation. The authors of \cite{Son:2009vu} followed a different but equivalent route, employing the
relation between the Kruskal extension of the Schwarzschild-AdS geometry and the Schwinger-Keldysh formalism \cite{Maldacena:2001kr,Herzog:2002pc}, together with the known connection between the latter and the generalized Langevin equation. These calculations were
later elaborated on in \cite{Giecold:2009cg,CasalderreySolana:2009rm,deboer2,Caceres:2010rm,Ebrahim:2010ra,CaronHuot:2011dr}.

In this paper, we generalize the original computation of \cite{brownian} to the case of non-commu\-ta\-ti\-ve Super-Yang-Mills (NCSYM). Non-commutative theories are known to lead to many qualitatively new phenomena, both classically and quantum mechanically; in particular, the existence of non-local interactions reminiscent of the UV/IR mixing found in string theory\cite{Matusis:2000jf}\footnote{For a review of non-commutative quantum field
theories, see for example \cite{Minwalla:1999px, Douglas:2001ba}.}. A simple model of non-commutativity was described by Bigatti and Susskind \cite{Bigatti:1999iz} where they considered a pair of opposite charges moving in a strong magnetic field. In the limit of large magnetic field, the charges are frozen into the lowest Landau levels and the interactions of such particles include the Moyal bracket characteristic of field theories on noncommutative space. Similarly, in the context of string theory, it was shown that the endpoints of open strings constrained to a D-brane in the presence of a constant Neveu-Schwarz $B$-field\footnote{By gauge invariance, this is equivalent to a constant magnetic field on the brane.} satisfy also the non-commutative algebra. A holographic realization of NCSYM was given a short time afterwards in \cite{Hashimoto:1999ut,Maldacena:1999mh}.

One of the main motivations to study Brownian motion in the presence of non-\-commuta\-ti\-vi\-ty is the idea that non-local interactions might lead to significant deviations in the behavior of the thermal properties of the theory \cite{Fischler:2000fv,Fischler:2000bp}. In \cite{Edalati:2012jj} for instance, it was found that the rate of decay of a fluctuation propagating in this thermal bath is remarkably larger than in the case of ordinary SYM, which leads to faster thermalization. Such a property is possibly linked to the connection between the ultraviolet and infrared regimes of the theory, which implies in particular that the transverse size of dipoles grows with their longitudinal momentum \cite{Bigatti:1999iz}. In order to investigate further properties of this non-commutative system, we study in this paper the holographic realization the Brownian motion of a heavy quark. More specifically, our aim is to formulate a Langevin equation that accounts for the effects of non-commutativity and to study diffusion processes within the plasma.

The structure of this paper is as follows: in section \ref{section2}, we describe the gravity dual to NCSYM presented in \cite{Hashimoto:1999ut,Maldacena:1999mh} and, following \cite{brownian}, we review the generalities of Brownian motion in the context of the gauge/gravity correspondence. In section \ref{secSYM} we study the case of Brownian motion in ordinary SYM with a magnetic field, which is achieved by the introduction of a gauge field in the open string sector\footnote{Here, we should emphasize that we do not expect to recover the non-commutative results in the strong $B$ limit. On one hand, there is a critical magnetic field at which Schwinger pair production is energetically favored \cite{Bolognesi:2012gr}. On the other hand, for large magnetic fields the backreaction on the geometry is unsuppressed.}. Apart from being an interesting problem in its own right\footnote{Indeed, many holographic systems present interesting features when a background magnetic field is turned on. For a recent review on this topic see \cite{Bergman:2012na}.}, this exercise helps us to gain some intuition and to set the grounds of our computations. In section \ref{sec4} we turn to the study of Brownian motion in NCSYM. We begin by postulating a Langevin equation for the non-commutative plasma similarly to the one describing the quark in the presence of a magnetic field. We show that this equation correctly captures behavior of Brownian particle. This equation is expressed naturally in terms of matrices and given its structure, it automatically implies that fluctuations along different directions are correlated. We then compute holographically the drag coefficient from the response of the Brownian particle to an external force and, finally, we study the diffusion process of the quark within the plasma, which turns out to be unaffected by non-commutativity. In section \ref{conclusions} we make some comments about our results and close with conclusions.

\section{Background and String Action\label{section2}}

\subsection{A gravity dual of NCSYM}

Quantum field theories on noncommutative spaces have been studied intensively in the last few years. The essential postulate of non-commutativity is the commutation relation
\be
\left[x^\mu,x^\nu\right]=i\theta^{\mu\nu},
\ee
where $\theta^{\mu\nu}$ is a real and antisymmetric rank-2 tensor. The algebra of functions on a non-commutative space can be viewed as an algebra of
ordinary functions with the product deformed to the non-commutative Moyal product, defined by
\be
\left(\phi_1\star\phi_2\right)(x)\equiv e^{\frac{i}{2}\theta^{\mu\nu}\partial_\mu^y\partial_\nu^z}\phi_1(y)\phi_2(z)\big|_{y=z=x}.
\ee

Non-commutative field theories emerge in string theory, as the worldvolume theory of D-branes with a constant Neveu-Schwarz $B$-field provided that one takes a special limit to decouple the open and closed string sectors \cite{Douglas:1997fm, Ardalan:1998ce,Seiberg:1999vs}. Basically, one scales the string tension to infinity and the closed string metric to zero while keeping the $B$-field fixed. For the discussion in the present paper, we will focus specifically on the four-dimensional $SU(N)$ NCSYM at finite temperature, taking the non-commutativity parameter to be non-zero only in the $(x^2,x^3)$-plane, that is $[x^2,x^3]\sim i \theta$. In the spirit of the gauge/gravity correspondence, the dynamics of this theory at large $N$ and at strong 't Hooft coupling should be described by a bulk gravitational system. Indeed, a gravity dual for this theory was proposed in \cite{Hashimoto:1999ut,Maldacena:1999mh} which, in the string frame, reads
\begin{eqnarray}\label{backg}
ds^2 &=& G_{mn}dx^mdx^n=\ap R^2 \left[u^2\left(-f(u) dx_0^2+dx_1^2+h(u)\left(dx_2^2+dx_3^2\right)\right)+ \frac{du^2}{u^2f(u)}+ d\mathrm{\Omega}_5^2\right]\nonumber\\
e^{2\Phi} &=& {\hat g}^2h(u)\nonumber\\
B_{23} &=& \ap R^2a^2u^4h(u)\\
C_{01} &=& \frac{\ap a^2R^2}{\hat g}u^4,\nonumber\\
\tilde F_{0123u} &=& \frac{4\ap ^2R^4}{{\hat g}}u^3h(u)\nonumber
\end{eqnarray}
where $R^4 = 4 \pi \hat g N=\lambda$ and $\hat g$ is the value of the string coupling in the IR. In the expressions above,
\be
f(u)=1-\frac{u_h^4}{u^4}
\ee
is the usual Schwarzschild factor and
\be
h(u)=\frac{1}{1+a^4u^4}.
\ee
The parameter $a$ which appears in the above expressions is related to the non-commutativity parameter $\theta$ of the boundary theory  through $a=\lambda^{1/4}\sqrt{\theta}$. Here, $\lambda$ is the 't Hooft coupling of the boundary large-$N$ NCSYM theory. The radial direction $u$ is mapped holographically into a energy scale in the gauge theory, in such a way that $u\rightarrow\infty$ and $u\rightarrow 0$ are respectively the UV and IR limits\footnote{Throughout this paper, we use the terms ``UV'' and ``IR'' with respect to the boundary energy. Then, in bulk terms, UV means near the boundary whereas IR means near the horizon.}. The directions $x^\mu\equiv(t,\vec{x})$ are parallel to the
boundary and are directly identified with the gauge theory directions. The five-sphere coordinates are associated with the global $SU(4)$ internal symmetry group, but they will play no role in our discussion.

The Hawking temperature of the above solution, which is interpreted as the temperature of the non-commutative boundary theory, is given by $T=u_h/\pi$. Notice that this temperature is the same as the temperature one obtains for the Schwarzschild-AdS$_5$ solution, which is dual to a thermal state of four-dimensional SYM. Indeed, it is easy to show that all the thermodynamic quantities obtained from \eqref{backg} are the same as the ones obtained from the Schwarzschild-AdS$_5$ solution \cite{Maldacena:1999mh}. In the limit of vanishing temperature ($u_h\rightarrow0$), we
are left with a geometry that is dual to the vacuum of the NCSYM. The closed string sector describing small or large fluctuations on top of this background captures the nonperturbative gluonic (+ adjoint scalar and fermionic)
physics. The black hole geometry (\ref{backg}) is a special case of such a fluctuation, and is dual to a thermal ensemble of the theory.

For $u \ll a^{-1}$,  the background \eqref{backg} goes over to the AdS$_5 \times {\rm S}^5$ solution.  This observation just reflects the fact that the non-commutative boundary theory goes over to the ordinary commutative SYM theory at length scales much greater than $\lambda^{1/4}\sqrt{\theta}$. On the other hand, for $u\gg a^{-1}$ the background \eqref{backg} shows significant deviation form the AdS$_5 \times {\rm S}^5$ solution and the bulk spacetime is no longer asymptotically AdS. The boundary theory interpretation of this regime just means that the effect of non-commutativity becomes pronounced for length scales which are at the order of, or smaller than, $\lambda^{1/4}\sqrt{\theta}$ \footnote{Since the reliability of the solution \eqref{backg} requires $R^4=\lambda$ to be large, the effect of non-commutativity in the boundary theory is visible even at large length scales \cite{Maldacena:1999mh}. This should be compared with the weak-coupling regime of the theory where the threshold length scale (beyond which the theory becomes effectively commutative) is roughly at the order of $\sqrt{\theta}$.}. To summarize, the background \eqref{backg} represents a flow in the boundary theory from a UV fixed point which is NCSYM at large $N$ and large $\lambda$ to an IR fixed point given by the ordinary commutative SYM (again, at large $N$ and large $\lambda$).

\subsection{Heavy quarks and holographic Brownian motion\label{holbrom}}
From the gauge theory perspective, the introduction of an open string sector
associated with a stack of $N_f$ D7-branes in the geometry (\ref{backg}) is equivalent to the
addition of $N_f$ hypermultiplets in the fundamental representation of the gauge group $SU(N)$, and these are the degrees
of freedom that we will refer to as quarks. For $N_f\ll N$, we can neglect
the backreaction of the D7-branes on the geometry; in the gauge theory perspective, this
 corresponds to working in a ``quenched'' approximation that ignores quark loops. For simplicity, we will take $N_f=1$.
The probe brane covers the four gauge theory directions $x^\mu$, and is spread along the
radial direction from $u\rightarrow\infty$ to $u=u_m$ where it ends smoothly\footnote{At this position the $\mathrm{S}^3\subset\mathrm{S}^5$ that it is wrapped on shrinks down to zero size.}.

An isolated quark is dual to an open string that extends radially from the flavor
brane at $u=u_m$ to the horizon at $u=u_h$\footnote{For a review of quark dynamics in the context of the gauge/gravity correspondence see \cite{Chernicoff:2011xv}.}. The string couples to both the metric and the $B$-field so its dynamics follows from the action
$S=S_{\text{NG}}+S_{\text{B}}$,\footnote{The string also couples to the dilaton but this coupling is suppressed by a factor of the string length.}
\be\label{nambugoto}
S\equiv\int_\Sigma d\tau d\sigma\, \cL = \frac{1}{2\pi\ap}\int_\Sigma d\tau d\sigma\left(\sqrt{-\det g_{ab}}+B_{mn}\p_\tau X^m\p_\sigma X^n\right)
\ee
where $g_{ab}=G_{mn}\p_aX^m\p_bX^n$ is the induced metric on the worldsheet $\Sigma$. We choose to work in the static gauge, where $\tau=t$, $\sigma=u$. One can easily verify that the embedding $X^m=(t,0,0,0,u)$ is a trivial solution and this correspond to a  quark that is is equilibrium in the thermal bath.

Notice that the string is being described in first-quantized language
and, as long as it is sufficiently heavy, we are allowed to treat it semiclassically.
In gauge theory language, we are coupling a first-quantized quark to the SYM fields, and then carrying out the full path integral over the strongly-coupled fields, treating the path integral over the quark trajectory in a saddle-point approximation. The mass of the quark $m$ is related to the position of the flavor brane $u_m$ and can be obtained by a straightforward computation:
\be\label{mass}
 m= {1\over 2\pi\ap}\int_{u_h}^{u_m} du\,\sqrt{g_{tt}\,g_{uu}}
 ={R^2\over 2\pi}(u_m-u_h)\approx{R^2\over 2\pi}u_m,\qquad\text{for}\quad u_m\gg u_h.
\ee

Now, we want to study fluctuations over the above embedding. In particular, we are interested in motion on the Moyal plane, so our ansatz for the perturbations will be $X^m=(t,0,X_2(t,u),X_3(t,u),u)$. Using this, the induced metric has the following components:
\bea
g_{tt}&=&\ap R^2u^2\left[-f+h\left(\dot{X}_2^2+\dot{X}_3^2\right)\right],\\
g_{uu}&=&\ap R^2u^2\left[h\left(X_2'^2+X_3'^2\right)+\frac{1}{u^4f}\right],\\
g_{tu}&=&\ap R^2u^2h\left(\dot{X}_2X'_2+\dot{X}_3X'_3\right),
\eea
where $\dot{X}_i\equiv\p_tX_i$ and $X_i'\equiv\p_uX_i$. Up to quadratic order in the perturbations, the action can be written as
\be\label{action}
S\approx\frac{R^2}{4\pi}\int dt du\left[u^4fh\left(X_2'^2+X_3'^2\right)-\frac{h}{f}\left(\dot{X}_2^2+\dot{X}_3^2\right)+2a^2u^4h\left(\dot{X}_2X_3'-\dot{X}_3X_2'\right)\right].
\ee
Note that we dropped the constant term that does not depend on $X_i$. We can consider also the situation in which one has a forced motion due to a electromagnetic field in the background.  This can be easily realized by turning on a world-volume $U(1)$ gauge field on the flavor brane. Since the
endpoint of the string is charged, this amounts to add the minimal coupling to the action $S=S_{\text{NG}}+S_{\text{B}}+S_{\text{EM}}$, where
\be\label{actem}
S_{\text{EM}}=\int_{\p\Sigma}\left(A_t+A_i \dot{X}_i\right) dt,
\ee
This will exert the desired force on our heavy quark. However, this coupling is just a boundary term, so it will not play any role for the string dynamics in the bulk, other than modify the boundary condition. We shall ignore this part of the action for now but we will come back to it later on.

Because $t$ is an isometry of the background (\ref{backg}), we can set
\be\label{fourier}
X(t,u)\sim e^{-i\omega t}g_\omega(u)
\ee
and use the frequency $\omega$ to label the basis of solutions to the equations of motion. Since the action (\ref{action}) is quadratic in the perturbations, we expect linear differential equations. The solutions are particularly easy to obtain near the horizon limit $u\sim u_h$, where the action reduces to
\bea
 S\!&\approx&\! u_h^2 h(u_h){R^2\over 4\pi}\int dt du_*
 \left[\left(X_2'^2+X_3'^2\right)-\left(\dot{X}_2^2+\dot{X}_3^2\right)+2a^2u_h^2\left(\dot{X}_2X_3'-\dot{X}_3X_2'\right)\right]\nonumber\\
 \!&\approx&\! u_h^2 h(u_h){R^2\over 4\pi}\int dt du_*
 \left[\left(X_2'^2+X_3'^2\right)-\left(\dot{X}_2^2+\dot{X}_3^2\right)\right].\label{action_NH}
\eea
Here, the primes denote derivatives with respect to the tortoise coordinate $u_*$, which is defined by
\be
 du_*={du\over u^2f(u)}.\label{tortoise}
\ee
Note also that the last term drops out because it is a total derivative. Thus, the equations of motion are then
\begin{align}
 (\partial_{u_*}^2-\partial_{t}^2)X_i= 0,\label{eom_NH}
\end{align}
which show that in this region $X_i$ behave like massless Klein-Gordon scalars in flat space. The two independent solutions are
\be
X^{(\text{out})}_i(u)=e^{-i\omega t}g_i^{(\text{out})}(u)\sim e^{-i\omega(t-u_*)}\label{solout}
\ee
and
\be
X^{(\text{in})}_i(u)=e^{-i\omega t}g_i^{(\text{in})}(u)\sim e^{-i\omega(t+u_*)},\label{solin}
\ee
corresponding to outgoing and ingoing waves respectively. Near the horizon one finds that
\begin{align}
 u_*\sim {1\over 4u_h}\log\left({u-u_h\over u_h}\right)
 \label{tortoise_NH}
\end{align}
up to an additive numerical constant, so
\be\label{normal}
g^{(\text{out/in})}(u)\sim\left(\frac{u}{u_h}-1\right)^{\pm i\omega/4u_h}.
\ee
Away from the horizon, these solutions will have a complicate dependence, but it still holds that $g^{(\text{out})}=g^{(\text{in})}\,\!^*$ (see appendix \ref{appsols}).

Standard quantization of quantum fields in curved spacetime
\cite{Birrell} leads to a mode expansion of the form
\be\label{modex}
 X_i(t,u)=
 \int_0^\infty {d\omega\over 2\pi}
 [a_{\omega} u_{\omega}(t,u)+a_{\omega}^\dagger u_{\omega}(t,u)^*],
\ee
where the functions $u_{\omega}(x)$ correspond to a basis with positive-frequency modes. These modes can be expressed as a linear combination of outgoing and ingoing waves with arbitrary coefficients, i.e.,
\be\label{defab}
 u_\omega(t,u)=
 A
 \left[g^{(\text{out})}(u)+B \, g^{(\text{in})}(u)\right]e^{-i\omega t}.
\ee
The constant $B$ is fixed through the boundary condition at $u=u_m$ but one generally obtains that it is a pure phase $B=e^{i\theta}$ (see sections \ref{hawk1} and \ref{hawk1}). The outgoing and ingoing modes have, then, the same amplitude and this implies that the black hole, which emit
Hawking radiation, can be in thermal equilibrium \cite{Hemming:2000as}. The constant $A$ on the other hand, is obtained by requiring the normalization of the modes through the standard Klein-Gordon inner product.

For any functions $f_i(t,u)$ and $g_j(t,u)$ satisfying the equations of motion, the Klein-Gordon inner product is defined by
\be\label{inner}
 (f_i,g_j)_\sigma=-{i\over 2\pi\ap}\int_\sigma \sqrt{\tilde g}\,
 n^\mu G_{ij}\, (f_i\partial_\mu g_j^*-\partial_\mu f_i\, g_j^*),
\ee
where $\sigma$ is a Cauchy surface in the $(t,u)$ part of the metric, $\tilde g$ is the induced metric on
$\sigma$ and $n^\mu$ is the future-pointing unit normal to $\sigma$. It can be shown that this inner product is independent of the choice of $\sigma$ \cite{Birrell}, but for simplicity we take it as a constant-$t$ surface.

We want to normalize $u_\omega$ using (\ref{inner}). However the main contribution to the integral comes from the IR region \cite{deboer2}, which in terms of the tortoise coordinate is just
\be
(f_i,g_j)_\sigma= -i\delta_{ij}{R^2 \over 2\pi}u_h^2h(u_h)\int_{u_*\sim-\infty}\!\!\!\!\!\!\!\!\!\!\!\! du_*
 (f_i\, \dot{g}_j^*- \dot{f}_i\, g_j^*).\label{inner_prod_NH}
\ee
Of course, there is a contribution to the inner product from regions away
from the horizon, but because the near-horizon region is
semi-infinite in the tortoise coordinate $u_*$, the normalization of solutions is
completely determined by this region.

After some algebra, we find that
\be
A=\sqrt{\frac{\pi}{\omega R^2u_h^2h(u_h)}},
\ee
so that $(u_\omega,u_\omega)=1$, ensuring that the canonical commutation relations are satisfied:
\begin{align}
  [a_{\omega},a_{\omega'}]=
  [a_{\omega}^\dagger,a_{\omega' }^\dagger]=0,\qquad
  [a_{\omega},a_{\omega'}^\dagger]=2\pi\delta(\omega-\omega').\label{CCR}
\end{align}

In the semiclassical approximation, the string modes are thermally excited by Hawking radiation emitted by the worldsheet horizon. In
particular, they satisfy the Bose-Einstein distribution:
\be
 \left\langle a_\omega^\dagger\, a_{\omega'}\right\rangle
 ={2\pi\delta{(\omega-\omega')}\over e^{\beta \omega}-1}.
\ee
Using this and the mode expansion given in (\ref{modex}), we can derive a general formula for the displacement squared of the Brownian particle.

First of all, let us identify the position of the heavy quark as the string endpoint at the boundary $u=u_m$, i.e.,
\be
 x_i(t)\equiv X_i(t,u_m)=
 \int_0^\infty {d\omega\over 2\pi}
 \left[a_\omega u_\omega(t,u_m) + a_\omega^\dagger u_\omega(t,u_m)^*\right].
\ee
Then, it follows that
\be
\left\langle x_i(t)x_i(0)\right\rangle=\int_0^\infty {d\omega d\omega'\over (2\pi)^2}\left[\langle a_\omega a_{\omega'}^\dagger\rangle u_\omega(t,u_m)u_{\omega'}(0,u_m)^*+\langle a_\omega^\dagger a_{\omega'}\rangle u_\omega(t,u_m)^*u_{\omega'}(0,u_m)\right].
\ee
This has an IR divergence that comes from the zero point energy, which exists even at zero temperature. To avoid this we simply
regularize it by implementing the normal ordering ${:\! a_\omega a_\omega^\dagger \!:} \equiv {:\! a_\omega^\dagger
a_\omega\! :}$, and after doing so we get\footnote{Another way to regularize it is to use the canonical correlator introduced as in \cite{brownian}. However, this does not change the late-time or low-frequency behavior of the correlator.}
\bea\label{xxcorr}
\left\langle:\! x_i(t)x_i(0)\! :\right\rangle&=&\int_0^\infty{d\omega\over 2\pi}\frac{1}{e^{\beta\omega}-1}\left[u_\omega(t,u_m)u_{\omega}(0,u_m)^*+u_\omega(t,u_m)^*u_{\omega}(0,u_m)\right],\nonumber\\
&=&\int_0^\infty{d\omega\over 2\pi}\frac{2|A|^2\cos(\omega t)}{e^{\beta\omega}-1}\left|g^{(\text{out})}(u_m)+B \, g^{(\text{in})}(u_m)\right|^2.
\eea
From this correlator, we can compute the displacement squared of the quark as:
\be\label{disquared}
s_i^2(t)\equiv\left\langle:\!\left[x_i(t)-x_i(0)\right]^2\!:\right\rangle=\frac{4}{R^2u_h^2h(u_h)}\int_0^\infty \frac{d\omega}{\omega}\frac{\sin^2\left(\tfrac{\omega t}{2}\right)}{e^{\beta\omega}-1}\left|g^{(\text{out})}(u_m)+B \, g^{(\text{in})}(u_m)\right|^2,
\ee
where we have replaced the explicit dependence on $A$. For future reference we also compute the general form of the momentum correlator,
\bea\label{ppcorr}
\left\langle:\! p_i(t)p_i(0)\! :\right\rangle&=&-m^2\partial_t^2\left\langle:\! x_i(t)x_i(0)\! :\right\rangle,\nonumber\\
&=&\int_0^\infty{d\omega\over 2\pi}\frac{2m^2\omega^2|A|^2\cos(\omega t)}{e^{\beta|\omega|}-1}\left|g^{(\text{out})}(u_m)+B \, g^{(\text{in})}(u_m)\right|^2.
\eea

\section{Brownian Motion in SYM with a Magnetic Field\label{secSYM}}

Non-commutative SYM theory has a dual interpretation in terms ordinary SYM with a large and constant magnetic field. We start by studying the Brownian motion in this second system which, aside from being an interesting problem in its own right, helps us to gain some intuition and to set the physical grounds of our computations. Although we find that there are some similarities between these two configurations, our final results show that there are some features that are qualitatively different.

\subsection{Langevin dynamics in the presence of a magnetic field\label{BroM}}

The problem of the Brownian motion of a charged particle in an external magnetic field was first investigated almost fifty years ago in the seminal papers \cite{BM1,BM2}. This is an old topic that has originated a lot of interest and it is of great importance in the description of diffusion and transport of plasmas and heavy ions. Nowadays, together with the free Brownian motion, it is widely used as a classic textbook
example of how transport properties and correlation functions should be computed in generic situations governed by the Langevin equation.

The discussion of this section will be around the field theory description of Brownian motion in the presence of a magnetic field. Later on, we will show how to realize this phenomenon at strong-coupling, in terms of a probe string living in a black hole background.

Let us consider the Langevin equation of a charged particle of mass $m$ and unit
charge $q$, in presence of a magnetic field $\mathbf{B}$:
\be\label{langeB}
\dot{\mathbf{p}}(t)=-\gamma_o\, \mathbf{p}(t)+\mathbf{v}(t)\times\!\mathbf{B}+\mathbf{R}(t),
\ee
where $\mathbf{p}(t)=m\,\mathbf{v}(t)$ is the momentum of the Brownian particle and $\mathbf{v}(t)=\dot{\mathbf{x}}(t)$ its velocity. The terms in the right-hand side of (\ref{langeB}) correspond to the friction, lorentz force and random force, respectively, and $\gamma_o$ is a constant called the friction coefficient. One can think of the particle as moving under the influence of the magnetic field, losing energy to the medium and at the same time, getting random kicks as modeled by the random force.

As a first approximation, we can assume that the random force is white noise, with the following averages:
\be
\langle R_i(t)\rangle=0,\qquad\langle R_i(t)R_j(t')\rangle=\kappa_o\delta_{ij}\delta(t-t'),
\ee
and where $\kappa_o$ is a constant which, due to the fluctuation-dissipation theorem, is related to the friction coefficient through
\begin{align}\label{flucdis}
 \gamma_o&={\kappa_o\over 2\,m\,T}.
\end{align}
This is due to the fact that the frictional and random forces have the same origin at the microscopic level, i.e., collisions with the particles of the thermal bath.

If the magnetic field $\mathbf{B}=B\hat{x}$ is pointing along the $x$-direction, we can write (\ref{langeB}) in the matrix form
\be\label{langeB2}
\dot{p}_i(t)=-\mathrm{\Lambda}_{ij}p_j(t)+R_i(t),
\ee
where
\be
\mathrm{\Lambda}_{ij}=\left(
                        \begin{array}{ccc}
                          \gamma_o & 0 & 0 \\
                          0 & \gamma_o & -\omega_o \\
                          0 & \omega_o & \gamma_o \\
                        \end{array}
                      \right).
\ee
Here $\omega_o=B/m$ denotes the Larmor frequency. Since the magnetic field is oriented along the $x$-direction,
it affects the motion in the $y$ and $z$ directions only. Fluctuations along the $x$-direction decouple and are unaffected by the presence of the magnetic field. We thus restrict our attention to the fluctuations in the transverse plane.

In order to decouple the remaining equations we have to diagonalize the the above matrix. The normal modes are $p_{\pm}=p_2\pm i p_3$ with corresponding eigenvalues $\lambda_{\pm}=\gamma_o\mp i \omega_o$. Thus, defining $x_\pm=x_2\pm i x_3$ and $R_{\pm}=R_2\pm i R_3$ we get
\be
    \dot{p}_{\pm}(t)=-\lambda_\pm p_{\pm}(t)+R_{\pm}(t),
\ee
whose formal solution is given by
\begin{equation}\label{momentum}
p_{\pm}(t)  =  e^{-\lambda_{\pm} t}p_\pm(0) + \int_0^t e^{-\lambda_{\pm} (t-t')}R_{\pm}(t')dt',
\end{equation}
\begin{equation}\label{momentumx}
x_\pm(t) - x_\pm(0)  =  {{1}\over{m \lambda_{\pm}}}\left( (1-e^{-\lambda{\pm} t}) p_\pm(0) +\left[
\int_0^t{R_\pm(t') dt'} -\int_0^t {e^{-\lambda (t-t')} R_{\pm}(t') dt'}\right] \right).
\end{equation}

Using the above, we can immediately obtain $x_2(t), x_3(t)$ by taking the real and
imaginary parts of $x_{\pm}(t)$. $p_2(t)$ and $p_3(t)$ can also be obtained in a similar way. In thermal equilibrium,
the two-point correlation functions of $p_2(t)$ and $p_3(t)$ are given by
\begin{eqnarray}\label{twopnt1}
\left \langle p_2(t) p_2(0) \right \rangle & = & \frac{\kappa_o}{2\gamma_o} e^{-\gamma_o t} \cos(\omega_o t),\\
\left \langle p_3(t) p_3(0) \right \rangle  & = & \left \langle p_2(t) p_2(t') \right \rangle,\label{twopnt2}\\
\left \langle p_2(t) p_3(0) \right \rangle & = & \frac{\kappa_o}{2\gamma_o} e^{-\gamma_o t} \sin(\omega_o t)\label{twopnt3}.
\end{eqnarray}
The relevant point here is that, unlike in the case of free Brownian motion, now the
components $p_2(t)$ and $p_3(t)$ are correlated. But, the autocorrelator
$\left\langle p_i(t)^2\right\rangle$ of each individual component has the same value $ \sim\kappa_o/2\gamma_o= mT$ as in the
case of zero magnetic field. This can be easily understood by recognizing that $\left\langle p_i(t)^2\right\rangle$ is
twice the kinetic energy multiplied by the mass of the particle and that this quantity does not change by application of a magnetic field.
Thus, the time scale associated to the energy loss due to drag, $t_{\text{relax}}\sim1/\gamma_o$, is then independent of the magnetic field.

Another important scale related to the Brownian motion of the quark is the time associated to diffusion. This can be derived by computing the two-point correlation function of $x_2(t)$ and $x_3(t)$, from which one can infer the following late-time behavior for the displacement squared:
\be\label{difct}
s_i^2(t)=\left\langle\left[x_i(t)-x_i(0)\right]^2\right\rangle\sim 2Dt,\qquad\text{for}\quad t\gg1/\gamma_o.
\ee
where $D$ is called the diffusion constant. In the presence of a magnetic field, the fluctuation-dissipation theorem leads to
\be
D=\frac{1}{2m^2}\frac{\kappa_o}{\gamma_o^2+\omega_o^2}=\frac{T}{m}\frac{\gamma_o}{\gamma_o^2+\omega_o^2},
\ee
which decreases with increasing magnetic field. Thus, the relevant time scale for correlations $\sim D$ is smaller in this case and this implies that the diffusion process is more efficient.

The Langevin equation, as presented in (\ref{langeB}), captures the essential properties of the stochastic processes in Brownian motion, but it fails to give a physically consistent picture for sufficiently short times $t$, in which the particle suffers only a few or no impacts. It is a general feature of any dynamical system that the dynamical coherence becomes predominant in short time scales, or at high frequencies. Thus, we are led to a natural extension of the Langevin equation in the form \cite{mori,kubo}
\be \label{genlangeB}
\dot{\mathbf{p}}(t)=-\int_{-\infty}^t dt'\, \gamma(t-t')\, \mathbf{p}(t')+\mathbf{v}(t)\times\!\mathbf{B}+\mathbf{R}(t)+\mathbf{E}(t),
\ee
where
\be\label{RR}
\langle R_i(t)\rangle=0,\qquad\langle R_i(t)R_j(t')\rangle=\kappa_{ij}(t-t').
\ee
The main difference between the generalized Langevin equation (\ref{genlangeB}) and the usual one (\ref{langeB}) is that the friction
depends now on the past history of the particle through $\gamma(t)$, called the memory kernel, and that the random forces at different times are not independent. Note that we have also included a fluctuating external force $\mathbf{E}(t)$ that can be applied to the system (e.g., an external electric field).

For a magnetic field pointing in the $x$-direction $\mathbf{B}=B\hat{x}$ and focusing on the transverse fluctuations, we get
\be
    \dot{p}_{\pm}(t)=-\int_{-\infty}^t dt'\,\gamma(t-t') p_{\pm}(t')\pm i \omega_o p_{\pm}(t)+R_{\pm}(t)+E_{\pm}(t),
\ee
where $p_{\pm}=p_2\pm i p_3$, $R_{\pm}=R_2\pm iR_3$, $E_{\pm}=E_2\pm i E_3$ and $\omega_o=B/m$.

In frequency domain, the above equation is simply\footnote{Causality imposes that $\gamma(t)=0$ for $t<0$ so $\gamma[\omega]$ in this expression denotes the Fourier-Laplace transform,
$$\gamma[\omega]=\int_{0}^\infty dt\, \gamma(t) \,e^{i\omega t},$$
while $p_{\pm}(\omega)$, $R_{\pm}(\omega)$, and $E_{\pm}(\omega)$ are Fourier transforms, e.g.,
$$p_{\pm}(\omega)=\int_{-\infty}^\infty dt\, p_{\pm}(t)\,e^{i\omega t}.$$}
\be\label{langefour}
 p_\pm(\omega)={R_\pm(\omega)+E_\pm(\omega)\over \gamma[\omega]-i\omega\mp i\omega_o},
\ee
and taking the statistical average of the same we get
\be
 \left\langle p_{\pm}(\omega)\right\rangle=
 \mu_\pm(\omega)E_\pm(\omega),\qquad\text{with}\qquad
 \mu_\pm(\omega)\equiv{1\over \gamma[\omega]-i\omega\mp i\omega_o}.
\ee
The quantity $\mu_\pm(\omega)$ is known as the admittance. Thus, we can then determine the
admittance $\mu_\pm(\omega)$, and thereby $\gamma[\omega]$, by
measuring the response $\left\langle p_\pm(\omega)\right\rangle$ to an external fluctuating force.  In
particular, if the external force is taken to be
\be
 E_\pm(t)=e^{-i\omega t}K_{\pm},
\ee
then $\left\langle p_\pm(t)\right\rangle$ is just
\be
 \left\langle p_\pm(t)\right\rangle=\mu_\pm(\omega) e^{-i\omega t}K_\pm=\mu_\pm(\omega) E_\pm(t).
\ee
For late times (or low frequencies) the generalized Langevin equation reduces to its non-local progenitor, and the timescales associated to the
decay of the two-point function of the momentum as well as the displacement squared are the same as discussed before.
In particular, note that
\be
\mu_\pm(0)\equiv{1\over \gamma_o\mp i\omega_o}
\ee
so
\be
\gamma_o=t^{-1}_{\text{relax}}=\mathbf{Re}\left(\frac{1}{\mu_\pm(0)}\right).
\ee

For a quantity $\cO$, the power spectrum $I_\cO(\omega)$ is defined as
\be
  I_\cO(\omega)\equiv\int_{-\infty}^\infty dt\left\langle\cO(t_0)\cO(t_0+t)\right\rangle e^{i\omega t},
 \label{pwrspctr_def}
\ee
and it is related to the two-point function through the Wiener-Khintchine theorem
\be\label{thm}
\left\langle\cO(\omega)\cO(\omega')\right\rangle=2\pi \delta(\omega+\omega')I_\cO(\omega).
\ee
For stationary systems (\ref{pwrspctr_def}) is independent of $t_0$, so one can set $t_0=0$ in such situations. When the external force is set to zero, from (\ref{langefour}) it follows that
\be
 p_\pm(\omega)={R_\pm(\omega)\over \gamma[\omega]-i\omega\mp i\omega_o}=\mu(\omega)R_\pm(\omega),
\ee
and, using (\ref{thm}) one gets that
\be
 I_{p_\pm}(\omega)=\left|\mu(\omega)\right|^2I_{R_\pm}(\omega).
\ee
Therefore, the random force correlator appearing in (\ref{RR}) can be evaluated as
\be\label{kapa}
\kappa_\pm(\omega)=I_{R_\pm}(\omega)=\frac{I_{p_\pm}(\omega)}{\left|\mu_\pm(\omega)\right|^2}.
\ee
This will be important in the next section, to check the validity of the fluctuation-dissipation theorem.

\subsection{Bulk dynamics and the drag coefficient\label{matchingB}}
We now turn to the holographic realization of Brownian motion in the presence of a magnetic field.
Let us begin by considering the action (\ref{action}), which in the commutative limit $a\rightarrow 0$ reduces to
\be
S\approx\frac{R^2}{4\pi}\int dt du\left[u^4f\left(X_2'^2+X_3'^2\right)-\frac{1}{f}\left(\dot{X}_2^2+\dot{X}_3^2\right)\right].
\ee
This action describes the dynamics of a string in Schwarzschild-AdS$_5$ and it is dual to a quark in ordinary SYM at finite temperature. We then turn on a gauge field in the flavor brane of the form
\be
\vec{A}=\frac{B}{2}\left(y\hat{z}-z\hat{y}\right),
\ee
thus getting the desired magnetic field $\vec{B}=B\hat{x}$. As explained before, this only appears as a boundary term (\ref{actem}), so it will not affect the bulk dynamics of the string. The equations of motion coming from the above action are:
\be\label{eomb}
0=f\p_u\left(u^4fX_i'\right)-\ddot{X}_i.
\ee
We now proceed by expanding $X_i$ in modes as in (\ref{fourier}), i.e.,
\be
X_i(t,u)=e^{-i\omega t}g_i(u).
\ee
Then the equations of motion (\ref{eomb}) can be written as
\be\label{eomdeboer}
0=g_i''(y)+\frac{4y^3}{y^4-1}g_i'(y)+\frac{\nu^2y^4}{(y^4-1)^2}g_i(y)
\ee
where we defined dimensionless quantities
\be
y=\frac{u}{u_h},\quad \nu=\frac{\omega}{u_h},
\ee
and where primes denote now derivatives with respect to $y$. The wave equation for the modes (\ref{eomdeboer}) is independent of the magnetic field and it is exactly the same as the equation considered in \cite{brownian} for $d=4$.

We need to find the solutions of the equation (\ref{eomdeboer}). In general, it is not possible to
do this analytically for an arbitrary frequencies $\nu$ and hence we employ a low
frequency approximation $\nu\ll 1$ by means of the so-called matching
technique.  Here, we only write down the final result, relegating the details of the computation to appendix \ref{appsols}. The two solution that correspond to outgoing and ingoing waves at the horizon behave asymptotically
as
\be\label{solsec}
g^{(\text{out/in})}(y)\sim \left(1\mp\frac{i}{8} \nu (\pi -\log(4))\right) \left(1+\frac{\nu^2}{2y^2}\right)\mp \frac{i\nu}{3y^3}+\mathcal{O}(1/y^4).
\ee

We now consider the forced motion of our Brownian particle due to a fixed external magnetic field and a fluctuating electric field. As mentioned in section \ref{holbrom}, this amounts to the addition of the boundary term (\ref{actem}) to the action (\ref{nambugoto}), which imposes a boundary condition of the form
\be
\mathrm{\Pi}^u_i\big|_{\p\Sigma}\equiv\frac{\p\cL}{\p X'_i}\bigg|_{\p\Sigma}=F_i.
\ee
Here, $F_i=-(F_{it}+F_{ij}\dot{X}^j)$ is the usual Lorentz force and
\be
F_{\mu\nu}\equiv\p_\mu A_\nu-\p_\nu A_\mu=\left(
                                            \begin{array}{cccc}
                                              0 & E_1 & E_2 & E_3 \\
                                              -E_1 & 0 & -B_3 & B_2 \\
                                              -E_2 & B_3 & 0 & -B_1 \\
                                              -E_3 & -B_2 & B_1 & 0 \\
                                            \end{array}
                                          \right).
\ee
Then, for a magnetic field pointing in the $x$-direction, $B_1=B$, and transverse electric fields $E_2=E_2(t)$ and $E_3=E_3(t)$ we find a set of boundary conditions that inevitably mix the fluctuations along the transverse directions:
\be\label{bcb}
\frac{R^2}{2\pi}u^4fX'_2-B\dot{X}_3\bigg|_{u=u_m}=E_2,\quad \frac{R^2}{2\pi}u^4fX'_3+B\dot{X}_2\bigg|_{u=u_m}=E_3,
\ee
where $u_m$ denotes the position of the flavor brane. Our goal is now to compute the thermal expectation value (or one-point function) of the momentum, and then extract the admittance.

The general solution for $X_i$ is the sum of ingoing and outgoing modes at the horizon $X_i=A_i^{(\text{out})}X^{(\text{out})}+A_i^{(\text{in})}X^{(\text{in})}$, where $X^{(\text{out/in})}=e^{-i\omega t}g^{(\text{out/in})}$. In the semiclassical approximation, outgoing modes are always thermally excited because of Hawking radiation, while the ingoing
modes can be arbitrary. However, because the radiation is random, the phase of $A_i^{(\text{out})}$ takes random
values and, on average $\langle A_i^{(\text{out})}\rangle=0$. Then, we can write $\langle X_i\rangle=\langle A_i^{(\text{in})}\rangle e^{-i\omega t}g^{(\text{in})}(u)$, where $g^{(\text{in})}(u)$ correspond to the normalized solution given by (\ref{solsec}). For the remaining part of this section we will denote $\langle A_i^{(\text{in})}\rangle=A_i$ and $g^{(\text{in})}=g$ for simplicity.

In the Brownian motion literature, it is customary to work in a circular basis when an external magnetic field is included. Thus, we define
\be
X_{\pm}\equiv X_2\pm i X_3=e^{-i\omega t}g_{\pm}(u)\quad\text{and}\quad E_{\pm}\equiv E_2\pm i E_3=e^{-i\omega t}K_{\pm}.
\ee
In fact, the equations (\ref{bcb}) decouple in this basis. After some algebra we get
\be\label{bcb2}
\frac{R^2}{2\pi}u^4fg'A_{\pm}\pm\omega B g A_{\pm}\bigg|_{u=u_m}=K_{\pm},
\ee
where $A_{\pm}=A_2\pm i A_3$. Inverting this relation we obtain
\be
A_{\pm}=\frac{K_{\pm}}{\frac{R^2}{2\pi}u^4fg'\pm\omega B g}\bigg|_{u=u_m},
\ee
from which we can read the average position of the heavy quark, $\left\langle x_{\pm}(t)\right\rangle\equiv\left\langle X_{\pm}(t,u_m)\right\rangle=e^{-i\omega t}g(u_m)A_{\pm}$,
\be
\left\langle x_{\pm}(t)\right\rangle=e^{-i\omega t}K_{\pm}\frac{g}{\frac{R^2}{2\pi}u^4fg'\pm\omega B g}\bigg|_{u=u_m},
\ee
and
\be
\left\langle p_{\pm}(t)\right\rangle =E_{\pm}(t)\frac{-i\omega m g}{\frac{R^2}{2\pi}u^4fg'\pm\omega B g}\bigg|_{u=u_m}.
\ee
This expression allows us to read the admittance,
\be\label{muB}
\mu_\pm(\omega) =\frac{-i\omega m g}{\frac{R^2}{2\pi}u^4fg'\pm\omega B g}\bigg|_{u=u_m}.
\ee
In the zero frequency limit, and for a heavy quark $u_m\ll u_h$, we obtain
\be
\mu_\pm(0)=\frac{2 m }{\pi\sqrt{\lambda}  T^2\pm 2 i B},
\ee
from which we can infer
\be
\gamma_o=\frac{\pi\sqrt{\lambda}  T^2}{2m} \quad\text{and}\quad\omega_o=\pm\frac{B}{m}.
\ee
As expected, the friction coefficient is not modified by the presence of the magnetic field, which is consistent with the fact that the magnetic field does not do work. Also, we find that
\be
t_{\text{relax}}=\frac{1}{\gamma_o}=\frac{2m}{\pi \sqrt{\lambda}  T^2}.
\ee
This temporal scale dominates the late-time decay of the one-point function of $\mathbf{p}(t)$, for a quark that transverses the plasma, in agreement with the previous works \cite{Herzog:2006gh,Gubser:2006bz}. The late-time\footnote{There should be a smooth crossover with the early-time regime or high frequency limit. See for example \cite{early}.} behavior is dominated by the low frequency limit of the generalized Langevin equation, case in which (\ref{genlangeB}) reduces to its nonlocal progenitor (\ref{langeB}). From (\ref{momentum}) we can thus infer that,
\be\label{exppb}
\left\langle p_\pm(t)\right\rangle\sim e^{-\gamma_o t}e^{\pm i\omega_o t}.
\ee
This is exactly what is expected for the Brownian motion of a charged particle in the presence of a magnetic field.
To obtain the thermal averages of $p_2(t)$ and $p_3(t)$ we can simply take the real and imaginary parts of (\ref{exppb}).

\subsection{Diffusion and the fluctuation-dissipation theorem\label{hawk1}}

The purpose of this section is to compute holographically the displacement squared of the heavy quark and to extract from it the diffusion constant $D$. The upshot of the computation is summarized in equation (\ref{disquared}), which is valid for an arbitrary background. However, the functions $f^{(\pm)}_\omega(u)$ as well as the details for the computation of the constant $B$ vary according to each situation.

Before proceeding with the direct calculation of this quantity, it is useful to understand the boundary
conditions we want to impose on the fields. Although we are interested in the world-sheet theory of the probe string, in the static gauge the induced metric on the string inherits the geometric characteristics of the spacetime background. This means that the usual rules for correlators in the gauge/gravity correspondence apply in our case.

First, we need to impose an UV cutoff in order to have a quark with finite mass. The natural place to impose the cutoff is given by the location of the flavor brane $u_m$, which can be related to the mass $m$ of the quark through (\ref{mass}). The mass of the quark is chosen to be dominant scale of the system, so usually one would push the cutoff up to the boundary $u_m\rightarrow\infty$ and choose normalizable boundary conditions for the modes.
However, in our case that would correspond to a infinitely massive quark and there would be no
Brownian motion. A Neumann boundary condition at $u=u_m$ also does not work in our case because we would go back to the case of free Brownian motion. Instead, we use a mixed boundary condition in which the external magnetic field is on but the fluctuating electric fields are turned off. According to (\ref{bcb}) this is,
\be
\frac{R^2}{2\pi}u^4fX'_2-B\dot{X}_3\bigg|_{u=u_m}=0,\quad \frac{R^2}{2\pi}u^4fX'_3+B\dot{X}_2\bigg|_{u=u_m}=0,
\ee
or in terms of the modes $X_\pm$,
\be\label{bcbc}
\frac{R^2}{2\pi}u^4fX'_{\pm}\pm i B \dot{X}_\pm\bigg|_{u=u_m}=0.
\ee
Recall that $X_\pm$ can be expressed as the sum of outgoing modes and ingoing modes found previously, with arbitrary coefficients. Following the convention introduced in (\ref{defab}), let us write
\be
 X_\pm(t,u)=A_\pm\left[g^{(\text{out})}(u)+B_\pm\, g^{(\text{in})}(u)\right]e^{-i\omega t}.
\ee
From (\ref{bcbc}) it follows then that
\be
B_\pm=-\frac{\frac{R^2}{2\pi}u^4fg^{(\text{out})}\,\!' \pm \omega B g^{(\text{out})}}{\frac{R^2}{2\pi}u^4fg^{(\text{in})}\,\!'\pm \omega B g^{(\text{in})}}\bigg|_{u=u_m}\equiv e^{i\theta_\pm}.
\ee
The fact that $B_\pm$ is a pure phase is self-evident, given that $g^{(\text{out})}=g^{(\text{in})\,\!^*}$. To leading order in frequency one finds that
\be
B_\pm=\frac{\pi  R^2 T^2 (y_m^4-1)\mp 2i B y_m^4}{\pi  R^2 T^2 (y_m^4-1)\pm 2 i B y_m^4}+\mathcal{O}(\nu)=\left(\frac{\pi  R^2 T^2 \mp 2i B}{\pi  R^2 T^2 \pm 2 i B}+\mathcal{O}(1/y_m^4)\right)+\mathcal{O}(\nu),
\ee
from which one gets
\be
\left|g^{(\text{out})}+B_\pm g^{(\text{in})}\right|^2=\frac{4 \pi ^2 R^4 T^4}{4 B^2+\pi ^2 R^4 T^4}+\mathcal{O}(\nu).
\ee
The late-time behavior of the displacement squared can be then inferred from the low-frequency limit of (\ref{disquared}), i.e.,
\be
s^2_{\pm}(t)=\frac{16 \sqrt{\lambda} T^3}{4 B^2+\pi^2 \lambda T^4}\int_{0}^\infty \!\!d\omega\frac{\sin^2\left(\frac{\omega t}{2}\right)}{\omega^2}\sim\frac{4 \pi \sqrt{\lambda} T^3}{4 B^2+\pi^2 \lambda T^4}t.
\ee
Thus, as expected, we find that the diffusion constant defined as in (\ref{difct}) is given by
\be
D=\frac{2 \pi \sqrt{\lambda} T^3}{4 B^2+\pi^2 \lambda T^4}=\frac{T}{m}\frac{\gamma_o}{\gamma_o^2+\omega_o^2}.
\ee

Finally, in order to give an explicit check of the fluctuation dissipation theorem (\ref{flucdis}) we compute the random force autocorrelation appearing in (\ref{kapa}) to extract the coefficient $\kappa_o$. From (\ref{ppcorr}) we can evaluate the two-point correlator of the momentum $p_\pm$ as follows:
\bea
\left\langle:\! p_\pm(t)p_\pm(0)\! :\right\rangle&=&\int_0^\infty{d\omega\over 2\pi}\frac{2m^2\omega^2|A|^2\cos(\omega t)}{e^{\beta\omega}-1}\left|g^{(\text{out})}(u_m)+B_\pm \, g^{(\text{in})}(u_m)\right|^2,\nonumber\\
&=&\int_{-\infty}^\infty{d\omega\over 2\pi}\frac{m^2}{\sqrt{\lambda}\pi T}\frac{\beta|\omega|}{e^{\beta|\omega|}-1}\left|g^{(\text{out})}(u_m)+B_\pm \, g^{(\text{in})}(u_m)\right|^2e^{-i\omega t}.
\eea
Therefore,
\be
I_{p_\pm}(\omega)=\frac{m^2}{\sqrt{\lambda}\pi T}\frac{\beta|\omega|}{e^{\beta|\omega|}-1}\left|g^{(\text{out})}(u_m)+B_\pm \, g^{(\text{in})}(u_m)\right|^2,
\ee
and combining this with (\ref{muB}) one finds that
\be
I_{R_\pm}(\omega)=\pi  \sqrt{\lambda } T^3+\cO(\omega).
\ee
This gives us precisely a coefficient $\kappa_o=\pi  \sqrt{\lambda } T^3$ which agrees with (\ref{flucdis}), providing and explicit check of the fluctuation-dissipation theorem in the presence of a magnetic field.

\section{Brownian Motion in NCSYM\label{sec4}}

We now turn our attention to the study of Brownian motion in the non-commutative setup. The main difference here is that the closed string sector is modified by the inclusion of an antisymmetric $B$-field. In this setup, and after the appropriate decoupling limit, the effective field theory is described by a gauge theory living in a noncommutative space. It is interesting to study the similarities and differences with our previous computation when the magnetic field was introduced in the open string sector.

\subsection{Langevin dynamics in the non-commutative plasma\label{BroM2}}

To start, let us postulate a generalized Langevin equation of a particle in a non-commutative thermal bath:
\be\label{langenc}
\dot{p}_i(t)=-\int_{-\infty}^t dt'\,\mathrm{\Gamma}_{ij}(t-t')p_j(t')+R_i(t)+E_i(t),
\ee
where
\be
\langle R_i(t)\rangle=0,\qquad\langle R_i(t)R_j(t')\rangle=\kappa_{ij}(t-t').
\ee
In this case, the $B$-field does not appear explicitly in the Langevin equation, though its effect should be somehow be present through the coefficients $\mathrm{\Gamma}$ and $\kappa$. We propose that in this case $\mathrm{\Gamma}$ is a matrix that encodes the effects of the non-commutativity. In particular, for non-commutativity in the $(x^2,x^3)$-plane, we propose to write
\be\label{matrix}
\mathrm{\mathrm{\Gamma}}_{ij}(t)=\left(
                        \begin{array}{ccc}
                          \gamma(t) & 0 & 0 \\
                          0 & \gamma(t) & -\mathrm{\Omega}(t) \\
                          0 & \mathrm{\Omega}(t) & \gamma(t) \\
                        \end{array}
                      \right).
\ee
In the low frequency limit, the above equation becomes local in time, allowing us to write
\be
\dot{p}_i(t)=-\mathrm{\Gamma}_{ij}p_j(t)+R_i(t)+E_i(t),
\ee
where $\gamma(t-t')=\gamma_o\delta(t-t')$, $\mathrm{\Omega}(t-t')=\mathrm{\Omega}_o\delta(t-t')$ and $\kappa(t-t')=\kappa_o\delta(t-t')$. Note that this has exactly the same structure of (\ref{langeB2}), with $\gamma_o$ being the usual friction coefficient and $\mathrm{\Omega}_o$ playing the role of the Larmor frequency. Furthermore, if the fluctuation-dissipation theorem applies, we expect that the relation (\ref{flucdis}) is also true for the present configuration. Solutions (\ref{momentum}) and (\ref{momentumx}) hold in the low frequency limit. This means that the two-point correlators (\ref{twopnt1})-(\ref{twopnt3}), as well as the diffusive behavior of the displacement squared (\ref{difct}), act in the exact same way, but now with new coefficients $\gamma_o$, $\mathrm{\Omega}_o$ and $\kappa_o$ which might depend on the noncommutative parameter $\theta$.

The fluctuations along the $x^1$-direction are unaffected by the presence of the non-commu\-ta\-ti\-vi\-ty. We thus restrict our attention to the fluctuations on the Moyal plane. These fluctuations can be decoupled by working in the circular basis $p_{\pm}=p_2\pm i p_3$, $R_{\pm}=R_2\pm i R_3$ and $E_{\pm}=E_2\pm i E_3$. The eigenvalues of (\ref{matrix}) are found to be $\lambda_{\pm}=\gamma\mp i\mathrm{\Omega}$ so we can rewrite (\ref{langenc}) as
\be
    \dot{p}_{\pm}(t)=-\int_{-\infty}^t dt'\,\lambda_\pm(t-t') p_{\pm}(t')+R_{\pm}(t)+E_{\pm}(t).
\ee
In frequency domain this equation can be written as
\be\label{langefournc}
 p_\pm(\omega)={R_\pm(\omega)+E_\pm(\omega)\over \lambda_\pm[\omega]-i\omega},
\ee
and taking the statistical average we obtain
\be
 \left\langle p_{\pm}(\omega)\right\rangle=
 \mu_\pm(\omega)E_\pm(\omega),\qquad\text{with}\qquad
 \mu_\pm(\omega)\equiv{1\over \lambda_{\pm}[\omega]-i\omega}.
\ee
Then, by measuring the response $\left\langle p_\pm(\omega)\right\rangle$ due to an external force we can determine the
admittance $\mu_\pm(\omega)$ and thereby $\mu_\pm[\omega]$. In
particular, if the external force is taken to be
\be
 E_\pm(t)=e^{-i\omega t}K_{\pm},
\ee
then
\be
 \left\langle p_\pm(t)\right\rangle=\mu_\pm(\omega) e^{-i\omega t}K_\pm=\mu_\pm(\omega) E_\pm(t).
\ee
With this at hand, one can take the real and imaginary parts of $\mu_\pm$ to obtain $\gamma$ and $\mathrm{\Omega}$ respectively. Note also that in the zero frequency limit
\be
\mu_\pm(0)\equiv\frac{1}{\lambda_\pm[0]}={1\over \gamma_o\mp i\mathrm{\Omega}_o}.
\ee
The analysis of the power spectrum and two-point functions is the same as the one done in section \ref{BroM}. In particular, the equation (\ref{kapa}) relating the random force correlations with the momentum correlations should apply in this case, which will be useful to check the validity of the fluctuation-dissipation theorem for the current setup.

\subsection{Bulk dynamics and the drag coefficient\label{matchingB2}}

For the action given by (\ref{action}) we can derive the following equations of motion:
\bea
0&=&\frac{f}{h}\p_u\left(u^4fhX_2'-a^2u^4h\dot{X}_3\right)-\p_t\left(\dot{X}_2-a^2u^4fX_3'\right),\\
0&=&\frac{f}{h}\p_u\left(u^4fhX_3'+a^2u^4h\dot{X}_2\right)-\p_t\left(\dot{X}_3+a^2u^4fX_2'\right).
\eea
The term with mixed derivatives cancels out in both cases and one ends up with
\bea
0&=&\frac{f}{h}\p_u\left(u^4fhX_2'\right)-4a^2u^3fh\dot{X}_3-\ddot{X}_2,\\
0&=&\frac{f}{h}\p_u\left(u^4fhX_3'\right)+4a^2u^3fh\dot{X}_2-\ddot{X}_3.
\eea
These are two coupled partial differential equations. We now proceed by expanding $X_i$ in modes by setting
\be
X_2(t,u)=e^{-i\omega t}g_2(u),\quad\text{and}\quad X_3(t,u)=e^{i(\varphi-\omega t)}g_3(u).
\ee
We have introduced a phase difference for reasons that will become clear below. The equations of motion for the modes are
\bea
0&=&\frac{f}{h}\p_u\left(u^4fhg_2'\right)+4i\omega a^2u^3fh g_3e^{i\varphi}+\omega^2g_2,\\
0&=&\frac{f}{h}\p_u\left(u^4fhg_3'\right)-4i\omega a^2u^3fh g_2e^{-i\varphi}+\omega^2g_3.
\eea
If we choose $e^{i\varphi}=\pm i$, or equivalently $\varphi=\pm\pi/2$, the two equations of motion turn out to be the same. This motivates to consider the linear combinations $X_{\pm}=X_2\pm iX_3=e^{-i\omega t}g_{\pm}(u)$, where
\be
g_{2}=\frac{g_{+}+g_{-}}{2}\quad\text{and}\quad g_{3}=\frac{g_{+}-g_{-}}{2i}.
\ee
Not surprisingly, this is completely equivalent with the circular basis introduced in section \ref{secSYM} for the case of Brownian motion in a magnetic field. In this basis the equations of motion decouple, and can be rewritten as:
\be\label{eom}
0=g_{\pm}''(y)+\frac{4(1+b^4)y^3}{(y^4-1)(1+b^4 y^4)}g_{\pm}'(y)+\left(\frac{\nu^2 y^4}{(y^4-1)^2}\pm\frac{4\nu b^2y^3}{(y^4-1)(1+b^4y^4)}\right)g_{\pm}(y)
\ee
where we have defined dimensionless quantities
\be
y\equiv\frac{u}{u_h},\quad \nu \equiv \frac{\omega}{u_h},\quad b \equiv au_h,
\ee
and the primes denote derivatives with respect to $y$. The normal modes $X_{\pm}$ correspond to fluctuations with circular polarization, rotating clockwise or counterclockwise, respectively.
%(check sign for each case)

Explicit solutions to the above equations can be found in appendix \ref{appsol2}. The final expressions for outgoing and ingoing modes are
\be\label{solrealo}
g^{(\text{out})}_{\pm}(y)\sim \frac{i b^2 \nu y}{b^2\mp i}\left(1-\frac{\nu^2}{2y^2}\right)+ \left(1-\frac{i b^2 \nu}{b^2\mp i}-\frac{1}{8} i \nu (\pi -\log(4))\right) \left(1-\frac{\nu^2}{6y^2}\right)+\mathcal{O}(1/y^3),
\ee
and
\be\label{solreal}
g^{(\text{in})}_{\pm}(y)\sim -\frac{i b^2 \nu y}{b^2\pm i}\left(1-\frac{\nu^2}{2y^2}\right)+ \left(1+\frac{i b^2 \nu}{b^2\pm i}+\frac{1}{8} i \nu (\pi -\log(4))\right) \left(1-\frac{\nu^2}{6y^2}\right)+\mathcal{O}(1/y^3),
\ee
respectively.

We now exert an external fluctuating force $\vec{E}(t)$ on the string endpoint by turning on an electric field $F_{ti}=E_i$ on the flavor brane. Variation of  the whole action implies the standard dynamics for all interior points of the string, but now with boundary condition
\be
\mathrm{\Pi}_i^u\big|_{\p\Sigma}\equiv \frac{\p \cL}{\p X_i'}=E_i,
\ee
where $E_i$ is the external force.

From (\ref{action}), it follows that
\be\label{bcnc1}
\frac{R^2}{2\pi}\left(u^4fhX_2'-a^2u^4h\dot{X}_3\right)\bigg|_{u=u_m}=E_2,\quad \frac{R^2}{2\pi}\left(u^4fhX_3'+a^2u^4h\dot{X}_2\right)\bigg|_{u=u_m}=E_3,
\ee
where $u_m$ denotes the position of the D7-brane. Our goal is to find the admittance of the system for which we need the one-point function of the momentum. Again, it is convenient to work in the circular basis, i.e. $X_{\pm}=X_2\pm iX_3$ and $E_{\pm}=E_2\pm iE_3$. The general solution for $X_{\pm}$ is the sum of outgoing and ingoing modes $X_{\pm}=A_{\pm}^{(\text{out})}X_{\pm}^{(\text{out})}+A_{\pm}^{(\text{in})}X_{\pm}^{(\text{in})}$. However, as discussed before, the phase of $A_{\pm}^{(\text{out})}$ takes random values and on average $\langle A_{\pm}^{(\text{out})}\rangle=0$. Then, we can write $\langle X_{\pm}\rangle=\langle A_{\pm}^{(\text{in})}\rangle e^{-i\omega t}g^{(\text{in})}_{\pm}(u)$ and $E_{\pm}=e^{-i\omega t}K_{\pm}$ but, for simplicity, we will denote $\langle A_\pm^{(\text{in})}\rangle=A_\pm$ and $g^{(\text{in})}_\pm=g_\pm$ in the remaining part of this section.
In this basis the boundary conditions decouple:
\be\label{bcnc}
\frac{R^2}{2\pi}\left(u^4fh A_{\pm} g_{\pm}'(u)\pm \omega a^2u^4h  A_{\pm} g_{\pm}(u)\right)\bigg|_{u=u_m}=K_{\pm},
\ee
or, in terms of the dimensionless quantities $y$, $\nu$ and $b$ as defined previously,
\be
 A_{\pm}=\frac{2\pi K_{\pm}}{R^2u_h^3}\frac{1+b^4y^4}{(y^4-1) g_{\pm}'(y)\pm \nu b^2y^4g_{\pm}(y)}\bigg|_{y=y_m}.
\ee
The average position of the heavy quark, $\left\langle x_{\pm}(t)\right\rangle\equiv\left\langle X_{\pm}(t,y_m)\right\rangle$ is then
\be
\left\langle x_{\pm}(t)\right\rangle =\frac{2\pi}{R^2u_h^3}\frac{(1+b^4y^4)g_{\pm}(y)}{(y^4-1)g_{\pm}'(y)\pm \nu b^2y^4 g_{\pm}(y)}\bigg|_{y=y_m}K_{\pm}e^{-i\omega t},
\ee
and
\be
\left\langle p_{\pm}(t)\right\rangle =\frac{-2\pi i \omega m}{R^2u_h^3}\frac{(1+b^4y^4)g_{\pm}(y)}{(y^4-1)g_{\pm}'(y)\pm \nu b^2y^4 g_{\pm}(y)}\bigg|_{y=y_m}F_{\pm}(t),
\ee
from which we can read the admittance,
\be\label{adnc}
\mu_\pm(\omega) =\frac{-2\pi i \omega m}{R^2u_h^3}\frac{(1+b^4y^4)g_{\pm}(y)}{(y^4-1)g_{\pm}'(y)\pm \nu b^2y^4 g_{\pm}(y)}\bigg|_{y=y_m}.
\ee
In the zero frequency limit and for large mass\footnote{In the large mass expansion, we set the non-commutative parameter to a fixed value and then take the limit $y_m\rightarrow\infty$.} we get,
\be
\mu_\pm(0)=\frac{2 m \left(1+b^4 y_m^4\right)}{b^2 R^2 \pi T^2}\frac{1\mp i b^2}{b^2 y_m^4\pm i}\sim\frac{2m}{R^2 \pi T^2}(1\mp i b^2)+\cO(1/y_m^4).
\ee
As we can see, the real part coincides with the expected value for the commutative case, but now there is an additional part that is imaginary (and independent of the temperature, given that $b=a u_h=a\pi T$). For a quark that transverses the plasma one can read the following evolution at late times:
\be\label{exppnc}
\left\langle p_\pm(t)\right\rangle\sim e^{-\gamma_o t}e^{\pm i\mathrm{\Omega}_o t},
\ee
where
\be
\gamma_o=\frac{\pi\sqrt{\lambda}  T^2}{2m(1+\pi^4\lambda\theta^2T^4)}\quad\text{and}\quad\mathrm{\Omega}_o=\frac{ \pi^3 \lambda\theta T^4}{2m(1+\pi^4\lambda\theta^2T^4)}.
\ee
The friction coefficient is modified by the presence of the non-commutativity, and in this case
\be
t_{\text{relax}}=\frac{1}{\gamma_o}=\frac{2m(1+\pi^4\lambda\theta^2T^4)}{\sqrt{\pi\lambda} T^2}.
\ee
This agrees with a previous computation of the drag force in the Maldacena-Russo background \cite{Matsuo:2006ws,Roy:2009sw} (in the non-relativistic regime) and implies that the non-commutative plasma is less viscous in comparison to the commutative one.

\subsection{Diffusion and the fluctuation-dissipation theorem\label{hawk2}}

Now we turn to the computation of the displacement squared in the Maldacena-Russo background. First of all, we need to understand the boundary conditions we want to impose on the fields. In this case, the effect of non-commutativity is already present in the background itself, so the free Brownian motion is realized by imposing a Neumann boundary condition at $u=u_m$. According to (\ref{bcnc1}), this means
\be
\frac{R^2}{2\pi}\left(u^4fhX_2'-a^2u^4h\dot{X}_3\right)\bigg|_{u=u_m}=0,\quad\frac{R^2}{2\pi}\left(u^4fhX_3'+a^2u^4h\dot{X}_2\right)\bigg|_{u=u_m}=0,
\ee
or simply
\be\label{bcnc2}
f X_\pm' \pm ia^2\dot{X}_\pm\bigg|_{u=u_m}=0,
\ee
where again, we defined $X_\pm=X_2\pm i X_3$. The general solution is then written as a linear combination of outgoing and ingoing modes,
\be
 X_\pm(t,u)=A_\pm\left[g_\pm^{(\text{out})}(u)+B_\pm\, g_\pm^{(\text{in})}(u)\right]e^{-i\omega t},
\ee
and from the boundary condition (\ref{bcnc2}) we get that
\be
B_\pm=-\frac{f g_\pm^{(\text{out})}\,\!' \pm \omega a^2 g_\pm^{(\text{out})}}{f g_\pm^{(\text{in})}\,\!'\pm \omega a^2 g_\pm^{(\text{in})}}\bigg|_{u=u_m}\equiv e^{i\theta_\pm}.
\ee
Here, $B_\pm$ is also a pure phase given that, in this case, it still holds that $g_\pm^{(\text{out})}=g_\pm^{(\text{in})\,\!^*}$. To leading order in frequency, we obtain
\be
B_\pm=-\frac{(1\mp i b^2)(1\pm i b^2 y_m^4)}{(1\pm i b^2)(1\mp i b^2 y_m^4)}+\mathcal{O}(\nu).
\ee
With this at hand, we can also compute
\be
\left|g_\pm^{(\text{out})}+B_\pm g_\pm^{(\text{in})}\right|^2=\left(\frac{4}{(1+b^4)}+\mathcal{O}(1/y_m^4)\right)+\mathcal{O}(\nu),
\ee
and finally, by taking the low-frequency limit of (\ref{disquared}) we compute the late-time behavior of the displacement squared:
\be
s^2_{\pm}(t)=\frac{16 }{\pi^2 \sqrt{\lambda} T}\int_{0}^\infty \!\!d\omega\frac{\sin^2\left(\frac{\omega t}{2}\right)}{\omega^2}\sim\frac{4}{\pi \sqrt{\lambda} T}t.
\ee
Note that the factor that depends on $b$ exactly cancels with the $h(u_h)$ term that appears in the normalization constant $A$. We find that, surprisingly, the diffusion constant is not affected by non-commutativity, but the relation with respect to $\gamma_o$ and $\mathrm{\Omega}_o$ is still the same
\be
D=\frac{2}{\pi \sqrt{\lambda} T}=\frac{T}{m}\frac{\gamma_o}{\gamma_o^2+\mathrm{\Omega}_o^2}.
\ee
This suggests that the fluctuation-dissipation theorem holds even in the presence of non-commutativity. In order to explicitly check this, we compute the random force correlator in to extract the coefficient $\kappa_o$. From (\ref{ppcorr}) it follows that
\be
\left\langle:\! p_\pm(t)p_\pm(0)\! :\right\rangle=\int_{-\infty}^\infty{d\omega\over 2\pi}\frac{m^2(1+b^4)}{\sqrt{\lambda}\pi T}\frac{\beta|\omega|}{e^{\beta|\omega|}-1}\left|g^{(\text{out})}(u_m)+B_\pm \, g^{(\text{in})}(u_m)\right|^2e^{-i\omega t},
\ee
and hence
\be
I_{p_\pm}(\omega)=\frac{m^2(1+b^4)}{\sqrt{\lambda}\pi T}\frac{\beta|\omega|}{e^{\beta|\omega|}-1}\left|g^{(\text{out})}(u_m)+B_\pm \, g^{(\text{in})}(u_m)\right|^2.
\ee
At leading order, we find through (\ref{kapa}) that
\be
\kappa_o=\frac{\pi  \sqrt{\lambda } T^3}{1+\pi^4\lambda\theta^2T^4}.
\ee
This agrees with (\ref{flucdis}), thus providing an explicit check of the fluctuation-dissipation theorem for the non-commutative plasma.

\section{Discussion\label{conclusions}}

In this paper we carried out an analytical study of the dynamics of a heavy quark in two strongly-coupled systems at finite temperature: SYM in the presence of a magnetic field and NCSYM. The former was realized by studying the fluctuations of a string living in an AdS black hole background and turning on a gauge field in the open string sector. The latter was achieved by replacing the background for one that incorporates the effects of non-commutativity through the introduction of an antisymmetric $B$-field in the closed string sector.

For both systems, we found that the Langevin equation that describes the dynamics of such a quark has matrix coefficients and this fact induces correlations along the relevant directions. This is in complete agreement with the classical theory of Brownian motion in a magnetic field \cite{BM1,BM2}. We then displayed the basic properties of these equations by computing holographically the admittance and the random force autocorrelator and we showed that these two quantities are related through the usual fluctuation-dissipation theorem. The existence of such theorem is due to the fact that, at the microscopic level, friction and random forces have the same origin, i.e., interactions with the degrees of freedom of the thermal bath. Finally, we studied the diffusion of the quark in both systems and we showed that, although the non-commutative plasma is less viscous, the late-time behavior of the displacement squared is unaffected by the non-commutativity.

As explained in the introduction, one of the main motivations that led to this work was to establish whether the fast thermalization found in \cite{Edalati:2012jj} holds for more general situations or not. An important difference between the approach of this paper and that of \cite{Edalati:2012jj} is that here we studied the non-commutative plasma with a local probe. In the previous work, on the other hand, we considered composite non-local operators that are obtained by smearing the usual gauge covariant
operators over open Wilson lines\footnote{In non-commutative field theories, this modification makes the operators gauge-invariant. In fact, this class of operators are known to couple to the linearized supergravity fields \cite{Das:2000ur,Liu:2001ps}.}. The fast thermalization and large decay rates of the modes are then possibly related to the non-local character of the probes\footnote{It is well known that the presence of the open Wilson lines dominate the UV behavior of the two (and higher) point functions of the gauge invariant operators \cite{Gross:2000ba}.}. It would be interesting to further explore this question by probing the theory with probes of different `size' and study the associated timescales for the approach to thermal equilibrium. Two interesting possibilities to consider are Wilson loops and entanglement entropy.

It is important to emphasize that all of our computations were performed in the low frequency limit of the theory, in which case the analytical computations were under control. Going beyond the hydrodynamical regime might also offer new insights but it requires a numerical approach. For example, in \cite{Brattan:2012uy} it was shown that a large class of holographic quantum liquids exhibit novel collective excitations that appear due to the presence of a magnetic field. At high frequency, the dominant peak in the spectral function is associated to sound mode similar to the zero sound mode in the collisionless regime of a Landau Fermi liquid. The study of Brownian motion within this regime is beyond the scope of this paper, but it is also left for future works.

In conclusion, the results obtained in this paper shed additional light on the thermal nature of non-commutative gauge theories and suggest future directions of research. The present study constitutes yet another illustration of the usefulness of the gauge/gravity correspondence.

\section*{Acknowledgements}

This material is based upon work supported by the National Science Foundation under Grant No. PHY-0969020 and by the Texas Cosmology Center. W.T.G. is also supported by a University of Texas fellowship. We are grateful to M. Shigemori for a clarification about the overall normalization of the solutions and for pointing out a useful reference.

\appendix

\section{Solutions for the string embedding\label{appsols}}

In this appendix we well derive explicitly the solutions to the equations of motion considered in sections \ref{matchingB} and \ref{matchingB2}. At low frequencies, the solutions can be obtained by means of the matching technique \cite{brownian,deboer2} (see also \cite{matching}). To find these solutions, consider three regimes: (A) the near horizon solution ($y\sim1$) for arbitrary $\nu$, (B) the solution for arbitrary $y$ but $\nu\ll1$, and (C) the asymptotic solution ($y\rightarrow\infty$) for arbitrary $\nu$. The idea is to find the approximate solutions for each of the three regimes, and to match these to leading order in $\nu$.

\subsection{Solution for the AdS-Schwarzschild black hole\label{appsol1}}

Here we will solve the equation (\ref{eomdeboer}), by considering the three regimes alluded above.

\textbf{(A)} In this regime we can focus on the equation as $y\rightarrow1$:
\be\label{eqnA}
0=g''(y)+\frac{1}{y-1}g'(y)+\frac{\nu^2}{16(y-1)^2}g(y).
\ee
We have dropped the subindex $i$ because the equations of motion are the same for both $i=2,3$. The general solution in this regime is
\be
g^{A}(y)=A^{(\text{out})}(y-1)^{i\nu/4}+A^{(\text{in})}(y-1)^{-i\nu/4},
\ee
where the coefficients $A^{(\text{out})}$ and $A^{(\text{in})}$ correspond to outgoing and and
ingoing modes respectively. Normalizing each solution according to (\ref{normal}) and expanding for low frequencies, we obtain
\be\label{solAb}
g^{A(\text{out/in})}(y)=(y-1)^{\pm i\nu/4}\sim 1\pm\frac{i\nu}{4}\log(y-1)+\cO(\nu^2).
\ee

\textbf{(B)} For the regime of low frequencies we proceed to expand the solution as a series of the form
\be\label{expb}
g^{B}(y)=g_0(y)+\nu g_1(y)+\nu^2 g_2(y)+...
\ee
The first function can be obtained analytically by solving the equation
\be
0=g_0''(y)+\frac{4y^3}{y^4-1}g_0'(y).
\ee
The general solution goes as follows:
\be\label{solB0b}
g_{0}(y)=B_1+B_2\left(\tan^{-1}(y)+\tanh^{-1}(y)\right),
\ee
where, in order to have a reliable expansion in frequencies, we have to assume that the constants $B_1$ and $B_2$ are independent of $\nu$. In order to find the appropriate coefficients $B_1$ and $B_2$ to obtain the ingoing and outgoing modes we expand around the horizon and match with the first term in (\ref{solAb}). After doing so we obtain $B^{(\text{out/in})}_1=1$ and $B_2^{(\text{out/in})}=0$, so
\be
g^{(\text{out/in})}_{0}(y)=1.
\ee
The equation for $g_1(y)$ turns out to be the same as for $g_0(y)$, but now the matching has to be done with the second term in (\ref{solAb}). At the end we get
\bea
g^{(\text{out/in})}_{1}(y)&=&\pm\frac{1}{8}\left((2+i)\pi+2i\log(2)\right)\mp\frac{i}{2}\left(\tan^{-1}(y)+\tanh^{-1}(y)\right),\\
&\sim&\pm\frac{i}{4}\log(y-1)\,\,\,\text{as }\, y\rightarrow1.\nonumber
\eea
Then, up to this order we can write
\be
g^{B(\text{out/in})}(y)=1\pm\frac{1}{8}\nu\left((2+i)\pi+2i\log(2)\right)\mp\frac{i}{2}\nu\left(\tan^{-1}(y)+\tanh^{-1}(y)\right)+\cO(\nu^2).
\ee
Asymptotically, these solutions behave as
\be\label{solBpb}
g^{B(\text{out/in})}(y)\sim\left[1\mp\frac{i}{8} \nu (\pi -\log(4))+\cO(\nu^2)\right]\mp\frac{1}{y^3}\left[\frac{i}{3}\nu+\cO(\nu^2)\right]+\cO(1/y^4).
\ee

\textbf{(C)} The general solution in region C can be found perturbatively, as a series expansion in $1/y$. The leading order terms go as follows:
\be
g^{C}(y)= C_1 \left[1+\frac{\nu^2}{2y^2}+\mathcal{O}(1/y^4)\right]+  \frac{C_2}{y^3}\left[1-\frac{\nu^2}{10y^2}+\mathcal{O}(1/y^4)\right].
\ee
Again, comparing this with (\ref{solBpb}), in the low frequency limit, we obtain
\be
C_1^{(\text{out/in})}=1\mp\frac{i}{8} \nu (\pi -\log(4)),\quad\text{and}\quad C_2^{(\text{out/in})}=\mp\frac{i}{3}\nu.
\ee
Thus, the \emph{normalized} asymptotic solutions for modes corresponding to outgoing and ingoing waves at the horizon, can be written as
\be\label{solrealb}
g^{C(\text{out/in})}(y)\sim \left(1\mp\frac{i}{8} \nu (\pi -\log(4))\right) \left(1+\frac{\nu^2}{2y^2}\right)\mp \frac{i\nu}{3y^3}+\mathcal{O}(1/y^4).
\ee
This agrees with the solutions reported in \cite{brownian}. Note in particular that $g^{(\text{out})}(y)=g^{(\text{in})}(y)^*$.

\subsection{Solution for the Maldacena-Russo background\label{appsol2}}

Next, we solve (\ref{eom}) following a similar procedure as the one used for the commutative case.

\textbf{(A)} In this regime we can focus on the equation as $y\rightarrow1$:
\be
0=g_{\pm}''(y)+\frac{1}{y-1}g_{\pm}'(y)+\frac{\nu^2}{16(y-1)^2}g_{\pm}(y),
\ee
which is equivalent to (\ref{eqnA}). As expected, the IR is not affected by the non-commutativity. The general solution in this regime is
\be
g_{\pm}^{A}(y)=A^{(\text{out})}_{\pm}(y-1)^{i\nu/4}+A^{(\text{in})}_{\pm}(y-1)^{-i\nu/4},
\ee
where the coefficients $A^{(\text{out})}_{\pm}$ and $A^{(\text{in})}_{\pm}$ correspond to outgoing and
ingoing modes respectively. Normalizing these solution and expanding for low frequencies, we obtain
\be\label{solA}
g_{\pm}^{A(\text{out/in})}(y)=(y-1)^{\pm i\nu/4}\sim 1\pm\frac{i\nu}{4}\log(y-1)+\cO(\nu^2).
\ee

\textbf{(B)} In this regime we start by expanding the solution as a series in $\nu$:
\be\label{exp}
g_{\pm}^{B}(y)=g_{0\pm}(y)+\nu g_{1\pm}(y)+\nu^2 g_{2\pm}(y)+...
\ee
The first function can be obtained analytically by solving the equation
\be
0=g_{0\pm}\,\!''(y)+\frac{4(1+b^4)y^3}{(y^4-1)(1+b^4 y^4)}g_{0\pm}\,\!'(y).
\ee
The general solution goes as follows:
\be\label{solB0}
g_{0\pm}(y)=B_{\pm1}+B_{\pm2}\left[b^4y-\tfrac{1}{2}(1+b^4)\left(\tan^{-1}(y)+\tanh^{-1}(y)\right)\right],
\ee
where, in order to have a reliable expansion in frequencies, we have to assume that the constants $B_{\pm1}$ and $B_{\pm2}$ are independent of $\nu$.
We then expand near the horizon and match the above solution with the first term in (\ref{solA}) to obtain the outgoing and ingoing modes. We find that $B^{(\text{out/in})}_{\pm1}=1$ and $B^{(\text{out/in})}_{\pm2}=0$, so
\be\label{normB0}
g^{(\text{out/in})}_{0\pm}(y)=1.
\ee

Plugging (\ref{exp}) into the equation of motion and using (\ref{normB0}) we can can derive the following equation for the next leading order term of the normalized modes:
\be\label{pert1}
0=g_{1\pm}\,\!''(y)+\frac{4 \left(1+b^4\right) y^3}{\left(y^4-1\right) \left(1+b^4 y^4\right)}g_{1\pm}\,\!'(y)\pm\frac{4 b^2 y^3}{\left(y^4-1\right) \left(1+b^4 y^4\right)}.
\ee
The solution to this equation is
\bea\label{solB1}
g_{1\pm}(y)\!&=&\!B_{\pm1}+B_{\pm2}\left[b^4y-\tfrac{1}{2}(1+b^4)\left(\tan^{-1}(y)+\tanh^{-1}(y)\right)\right]\\
&&\!\mp\frac{1}{2b^2}\left(\tan^{-1}(y)+\tanh^{-1}(y)\right),\nonumber
\eea
and after expanding around the horizon and matching with the second term in (\ref{solA}) we obtain
\be
B^{(\text{out})}_{\pm1}=\pm\frac{(1-2 i) \pi \pm b^2 ((2+i) \pi -i (8-\log(4)))+\log(4)}{8(b^2\mp i)},\quad B^{(\text{out})}_{\pm2}=\frac{i}{b^2 (b^2\mp i)},
\ee
and
\be
B^{(\text{in})}_{\pm1}=\pm\frac{(1-2 i) \pi \mp b^2 ((2+i) \pi -i (8-\log(4)))+\log(4)}{8 (b^2\pm i)},\quad B^{(\text{in})}_{\pm2}=-\frac{i}{b^2 (b^2\pm i)}.
\ee
Asymptotically, the solutions $g^{B(\text{out/in})}_{\pm}(y)=g^{(\text{out/in})}_{0\pm}(y)+\nu g^{(\text{out/in})}_{1\pm}(y)+\mathcal{O}(\nu^2)$ behave as
\be\label{solBpout}
g^{B(\text{out})}_{\pm}(y)\sim\left[\frac{i b^2 \nu}{b^2\mp i}y+\left(1-\frac{i b^2 \nu}{b^2\mp i}-\frac{1}{8} i \nu (\pi -\log(4))\right)+\cO(\nu^2)\right]+\cO(1/y),
\ee
and
\be\label{solBpin}
g^{B(\text{in})}_{\pm}(y)\sim\left[-\frac{i b^2 \nu}{b^2\pm i}y+\left(1+\frac{i b^2 \nu}{b^2\pm i}+\frac{1}{8} i \nu (\pi -\log(4))\right)+\cO(\nu^2)\right]+\cO(1/y).
\ee

\textbf{(C)} The general solution in region C can be found as a series expansion in $1/y$. The leading order terms are:
\be
g^{C}_{\pm}(y)= C_{\pm1} y\left[1-\frac{\nu^2}{2y^2}+\mathcal{O}(1/y^3)\right]+ C_{\pm2} \left[1-\frac{\nu^2}{6y^2}+\mathcal{O}(1/y^3)\right].
\ee
Comparing this with (\ref{solBpout}) and (\ref{solBpin}) we obtain
\be
C^{(\text{out})}_{\pm1}=\frac{i b^2 \nu}{b^2\mp i},\quad C^{(\text{out})}_{\pm2}=1-\frac{i b^2 \nu}{b^2\mp i}-\frac{1}{8} i \nu (\pi -\log(4)),
\ee
and
\be
C^{(\text{in})}_{\pm1}=-\frac{i b^2 \nu}{b^2\pm i},\quad C^{(\text{in})}_{\pm2}=1+\frac{i b^2 \nu}{b^2\pm i}+\frac{1}{8} i \nu (\pi -\log(4)).
\ee
Thus, the \emph{normalized} asymptotic solutions, in the low frequency limit, and for modes corresponding to outgoing and ingoing waves at the horizon, can be written as
\be
g^{C(\text{out})}_{\pm}(y)\sim \frac{i b^2 \nu y}{b^2\mp i}\left(1-\frac{\nu^2}{2y^2}\right)+ \left(1-\frac{i b^2 \nu}{b^2\mp i}-\frac{1}{8} i \nu (\pi -\log(4))\right) \left(1-\frac{\nu^2}{6y^2}\right)+\mathcal{O}(1/y^3),
\ee
and
\be
g^{C(\text{in})}_{\pm}(y)\sim -\frac{i b^2 \nu y}{b^2\pm i}\left(1-\frac{\nu^2}{2y^2}\right)+ \left(1+\frac{i b^2 \nu}{b^2\pm i}+\frac{1}{8} i \nu (\pi -\log(4))\right) \left(1-\frac{\nu^2}{6y^2}\right)+\mathcal{O}(1/y^3).
\ee
It is clear that $g^{(\text{out})}_{\pm}(y)=g^{(\text{in})}_{\pm}(y)^*$ as was found for the case analyzed in appendix \ref{appsol1}.

\end{document}